\documentclass[12pt]{article}

\newcommand{\be}{\begin{equation}}
\newcommand{\ee}{\end{equation}}
\newcommand{\bi}[1]{\vspace{-3mm} \bibitem{#1}}
\usepackage{epsfig,amsmath,amssymb,graphics,graphicx} 

\usepackage{graphics,graphicx}
\usepackage{amssymb,amsmath}
\usepackage{amscd}
\usepackage{pb-diagram}
\usepackage{hyperref}
\usepackage{amsfonts}
\usepackage{epsfig}

\begin{document}

\begin{center}

{\it Discontinuity, Nonlinearity, and Complexity.
Vol.4. No.1. (2015) 11-23.}
\vskip 3mm

{\bf \large Lattice Model with Nearest-Neighbor and 
\vskip 3mm
Next-Nearest-Neighbor Interactions for Gradient Elasticity} \\

\vskip 7mm
{\bf \large Vasily E. Tarasov} \\
\vskip 3mm

{\it Skobeltsyn Institute of Nuclear Physics,\\ 
Lomonosov Moscow State University, Moscow 119991, Russia} \\
{E-mail: tarasov@theory.sinp.msu.ru} \\

\begin{abstract}
Lattice models for the second-order strain-gradient models 
of elasticity theory are discussed.
To combine the advantageous properties of two classes
of second-gradient models, 
we suggest a new lattice model that can be considered 
as a discrete microstructural basis for gradient continuum models. 
It was proved that two classes of the second-gradient models 
(with positive and negative sign in front the gradient) 
can have a general lattice model as a microstructural basis. 
To obtain the second-gradient continuum models we consider 
a lattice model with the nearest-neighbor 
and next-nearest-neighbor interactions 
with two different coupling constants.
The suggested lattice model gives unified description of 
the second-gradient models 
with positive and negative signs of the strain gradient terms. 
The sign in front the gradient is determined by 
the relation of the coupling constants of the nearest-neighbor 
and next-nearest-neighbor interactions.
\end{abstract}

\end{center}



\section{Introduction}

Elastic deformations of materials can be described by 
a microscopic approach based on lattice equations 
\cite{Born,Kosevich} 
and by a macroscopic approach based on continuum equations 
\cite{Hahn,Landau}.
Continuum equations for elasticity can be considered 
as a limit of lattice dynamics, where 
the length scales of infinitesimal continuum elements 
are much greater than that of 
inter-particle distances in the lattice \cite{TarasovSpringer}. 
The theory of nonlocal continuum mechanics was formally initiated 
by the papers of Eringen \cite{Eringer1972}.
Nonlocal elasticity theory is based on the assumption that 
the forces between material points are a long-range type, 
thus reflecting the long-range character 
of inter-atomic forces \cite{Eringer2002}. 
Nearest-neighbor and next-nearest-neighbor interactions 
for lattice particles are important type of 
interactions for materials with non-local properties. 
Non-local continuum mechanics can be considered by
two different approaches \cite{Rogula}:
the gradient elasticity theory (weak non-locality) and 
the integral elasticity theory (strong non-locality).
This article focuses on gradient models 
of non-local elasticity suggested in
\cite{Mindlin1964,Mindlin1965,Mindlin1968,Eringen1983,Aifantis1992}. 
Mindlin \cite{Mindlin1964,Mindlin1965,Mindlin1968} presented a 
theory of elasticity for materials with microstructure, 
where it was proposed to distinguish quantities of the microscale 
and the macroscale to take into account a weak non-locality.
Eringen \cite{Eringen1983} also formulated a theory of nonlocal elasticity, where the integrals are replaced by gradients.
Aifantis \cite{Aifantis1992} 
suggested to extend the linear elastic constitutive relations 
with the Laplacian of the strain.
Usually distinguish two following classes of gradient models 
with different signs of the strain gradient terms. 

The first class is formed by Laplacian-based 
gradient elasticity models 
that are described by the linear constitutive relations 
\begin{equation} \label{H-1p}
\sigma_{ij} = C_{ijkl} \, \varepsilon_{kl} 
+ l^2 \, C_{ijkl} \, \Delta \, \varepsilon_{kl} ,
\end{equation}
where $\varepsilon_{ij}$ is the strain, 
$\sigma_{ij}$ is the stress,
$C_{ikjl}$ is the stiffness tensor,
$\Delta$ is the Laplace operator 
and $l$ is the scale parameter. 
For $l^2=0$, we have the classical case of the linear elastic 
constitutive relations that is called the Hooke's law.
The main motivation for using the gradient elasticity of the form (\ref{H-1p})
has been the description of dispersive wave propagation through heterogeneous media. 
For the second-gradient models that are defined by (\ref{H-1p}), 
it is found that the model becomes unstable for a limited number of wave lengths, 
while in dynamics, instabilities are encountered for all shorter wave lengths. 
The corresponding equation for the displacements 
is unstable for wave numbers $k> 1/l^2$.

The second class of Laplacian-based gradient elasticity models 
is described by the linear elastic constitutive relations of the form
\begin{equation} \label{H-1n}
\sigma_{ij} = C_{ijkl} \, \varepsilon_{kl} 
- l^2 \, C_{ijkl} \, \Delta \, \varepsilon_{kl}  .
\end{equation}
This equation has the format of equation (\ref{H-1p})
although the sign of the higher-order term tends to be negative, 
not positive as in equation  (\ref{H-1p}). 
Note that whereas the strain gradients with positive sign are destabilizing,
the strain gradients with negative sign are stable. 
The strain gradients in equation (\ref{H-1n}) 
are equivalent to those derived from the positive-definite
deformation energy density, 
and therefore the strain gradients in equation (\ref{H-1n}) are stable. 
The opposite sign of the strain gradient term 
in equation (\ref{H-1p}) makes this term destabilizing. 
Instabilities manifest themselves in dynamics 
by an unbounded growth
of the response in time without external work. 
Instabilities are also related to loss of uniqueness in 
static boundary value problems.

The lattice models are very important in the elasticity theory \cite{Born,Kosevich}.  
At the same time, it is formed the opinion that 
the continuum models described by equation (\ref{H-1n}) 
cannot be obtained from lattice models.
It is usually assumed that the second class of the gradient models does not 
have a direct relationship with discrete microstructure and lattice models. 
This opinion is based on the properties
of the Taylor series that is used in homogenization procedure.
This problem is described in Section 2 of this paper.
In next sections, we propose a lattice model that allows us 
to remove the lack of the second class of models.
Moreover the suggested type of interaction allows us 
to have united approach to describe lattice models of
the strain-gradient elasticity of two classes. 
The suggested lattice models give unified 
description of the second-gradient models 
with positive and negative signs of the strain gradient terms. 
A feature of suggested approach is the existence of an operation that transforms 
\cite{JMP2006,JPA2006,TarasovSpringer} (see also \cite{Chaos2006,CNSNS2006})
the set of equations for coupled individual particles of lattice
into the equation of non-local continuum.

\section{Homogenization procedure and Taylor series approach}

In order to keep this paper self-contained, 
we briefly reproduce a derivation of 
the continuum equation for the gradient elasticity 
by homogenization approach \cite{Mindlin1968,MA2002,AA2011}.

In this section, the strain gradient models (\ref{H-1p}) 
are derived by means of homogenization 
of the displacement field of a discrete model.
In the lattice model, the particles 
are replaced by individual masses.
The interactions of particle are modeled by springs that connected the point masses.
For simplicity, it is assumed that 
we have one-dimensional lattice, where 
all particles have the same spring stiffness $K$,
the particle mass $M$ and the inter-particle distance $d$. 

The gradient elasticity models (\ref{H-1p}) have been derived
from the continualization of the response of a lattice. 
To illustrate this approach, we will consider the one-dimensional
system of particles and springs. 
All particles have mass $M$ and all springs have spring stiffness $K$.
The equation of motion of the particle $n$ is 
\begin{equation} \label{eq4}
M D^2_t u_n(t) = K \cdot (u_{n+1} -2 u_n(t) + u_{n-1}(t) ) 
+ F(n) ,
\end{equation}
where $M$ and $K$ are the particle mass and 
the spring stiffness, respectively, both of which are assumed to be uniform.

In the homogenization procedure, it is assumed the continuous displacement $u(x,t)$
equals to the lattice displacement $u_n(t)$ at particle $n$ by $u_n(t)=u(nd,t)$,
where the particle spacing is denoted as $d$. 
The displacement at the neighboring particles is found by means of a Taylor series as
\begin{equation} \label{Taylor}
u(x \pm d,t) = u(x,t) \pm d \, D^1_x u(x) + \frac{d^2}{2} \, D^2_x u(x) 
\pm \frac{d^3}{6} \, D^3_x u(x) + \frac{d^4}{24} \, D^4_x u(x) + O(d^5) .
\end{equation}
Next, the displacements of the lattice medium $u_{n \pm 1}(t)$ are expressed in terms 
of the continuous displacement.
These terms are substituted into equation (\ref{eq4}). 
After division by the cross-section area of the medium $A$ and the inter-particle distance $d$
it is found that
\begin{equation} \label{eq6}
\rho \, D^2_t u(x,t) = E \Bigl(  D^2_x u(x,t) +  \frac{d^2}{12} \, D^4_x u(x,t) \Bigr) + f(x) 
\end{equation}
with the mass density $\rho= M/ A d$, the Young's modulus $E=K d/A$, and $f(x)=F(x)/Ad$.
Note that all odd derivatives of $u$ have cancelled
and signs in front of second-order and fourth-order derivatives coincide.
This equation can be written as
\begin{equation} \label{eq6b}
D^2_t u(x,t) = C^2_e \, D^2_x u(x,t) +  \frac{d^2 C^2_e}{12} \, D^4_x u(x,t) +\frac{1}{\rho} f(x), 
\end{equation}
where $C_e=\sqrt{E/\rho}$ is the elastic bar velocity. 

When the kinematic relation $\varepsilon =D^1_x u$ is used, and
the equation of motion of the continuum is expresses as
\begin{equation} \label{rho}
\rho \, D^2_t u(x,t)= D^1_x \sigma (x,t) + f(x), \end{equation}
the constitutive relation can be retrieved as
\begin{equation} \label{sign-pos}
\sigma =E \Bigl( \varepsilon + \frac{d^2}{12} D^2_x \varepsilon \Bigr) .
\end{equation}
The positive sign in the relation (\ref{sign-pos})
follows directly from the positive sign of $d^4$-term
in the Taylor series (\ref{Taylor}).

The second-gradient term is preceded by a positive sign, where $d=l \sqrt{12}$.
This procedure illustrates the close relation between the discrete microstructure 
and the gradient non-local continuum
with positive sign in (\ref{H-1p}) and (\ref{sign-pos}). 

The homogenization procedure as shown above uniquely leads to a second-order
strain gradient term that is preceded by a positive sign.
The second-gradient model with negative sign
\begin{equation} \label{Minus-1}
\sigma =E \Bigl( \varepsilon - l^2 \, D^2_x \varepsilon \Bigr) 
\end{equation}
cannot be derived from a microstructure of 
lattice particles by this homogenization procedure.


\section{Equations of motion for lattice particles}

In this paper, we shall use a simplest model 
to describe the lattice vibration, 
where all particles are displaced in one direction. 
We also assume that the displacement of particle from its equilibrium position 
is determined by a scalar rather than a vector. 
This model allows us to describe the main properties 
of a vibrating  lattice by using simple equations. 

Let us consider a one-dimensional lattice system of 
interacting particles that are described by the equations of motion
\begin{equation} \label{Main_Eq}
M \frac{d^2 u_n(t)}{d t^2} = 
g_2 \, \sum_{{m=-\infty}}^{+\infty} \; K_2(n,m) \; u_m(t)
+ g_4 \, \sum_{{m=-\infty}}^{+\infty} \; K_4(n,m) \; u_m(t) + F (n) ,
\end{equation}
where $M$ is the mass of particle, 
$u_n(t)$ are displacements from the equilibrium, 
$g_2$ and $g_4$ are coupling constants. 
The terms $F(n)$ characterize an interaction of the particles   
with the external on-site force. 
Let us note some properties of the coefficients $K_s(n, m)$. 
If we assume the lattice to be displaced as a whole: 
$u_n(t) = u =\operatorname{const}$, then 
the internal lattice state cannot be changed 
in case of absence of external forces $F (n)=0$.
As a result, equations (\ref{Main_Eq}) give 
\begin{equation} \label{Ko21}
\sum_{{m=-\infty}}^{+\infty} \; K_s (n,m) =
\sum_{{m=-\infty}}^{+\infty} \; K_s (m,n) = 0 , \quad (s=2;4)
\end{equation}
for all $n$. 
This requirement is well-known and it follows from the conservation of total momentum in the lattice \cite{Kosevich}.

It seems that two terms of (\ref{Main_Eq}) with 
$K_2(n,m)$ and $K_4(n,m)$
can be combined into one without loss of generality.
These terms are presented separately to use 
two different interaction constant $g_2$ and $g_4$ 
for two type of interactions such as
the nearest-neighbor and next-nearest-neighbor interactions. 
The advantage of such representation is manifested 
in Section 5. 
Using two different coupling constants allows us 
to derive the constitutive relation (\ref{H-1p}) and (\ref{H-1n})
with positive and negative 
depending on the relative values 
of the coupling constants $g_2$ and $g_4$.

For an unbounded homogeneous lattice, 
due to its homogeneity the matrix 
$K_s(n, m)$ has the form 
$K_s(n, m) = K_s(n - m)$ where $s=2;4$.
In a simple lattice each particle is an inversion center, and then we have 
\[ K_s(n-m) = K_s(m-n)= K_s(|n-m|) . \]
Using the condition (\ref{Ko21}), 
we can represent equations (\ref{Main_Eq}) in the form
\begin{equation} \label{Main_Eq2}
M \frac{d^2 u_n(t)}{d t^2} = 
g_2 \, \sum_{{m=-\infty}}^{+\infty} \; 
K_2(n-m) \; \Bigl( u_n-u_m \Bigr) + 
g_4 \, \sum_{{m=-\infty}}^{+\infty} \; 
K_4(n-m) \; \Bigl( u_n-u_m \Bigr) + F (n) .
\end{equation}
These equations of motion take into account the translation invariance condition  
of a lattice structure with respect to its displacement as a whole. 
In equation (\ref{Main_Eq2}) the interaction terms are translation invariant. 
The non-invariant interaction terms lead to the divergences 
in the continuous limit \cite{TarasovSpringer}.

In this paper we consider the lattice model (\ref{Main_Eq2}) 
with the nearest-neighbor and next-nearest-neighbor interactions only.
Then the first term $K_2(n-m)$ describes 
the nearest-neighbor interaction by
\begin{equation} \label{Knm2}
K_{2}(n-m) = - \Bigl( \delta_{n-m,1} + \delta_{m-n,1}\Bigr) 
\end{equation}
with the coupling constant $g_2$. 
The second term $K_4(n-m)$ describes 
the next-nearest-neighbor interaction by
\begin{equation} \label{Knm4}
K_{4}(n-m) = - \Bigl( \delta_{n-m,2} + \delta_{m-n,2}\Bigr) ,
\end{equation}
and $g_4$ is the coupling constant of 
this type of interaction.

We can give a discrete mass-spring system that corresponds 
to the suggested lattice model (\ref{Main_Eq2}).
In Figure 1, we present the mass-spring system
with the nearest-neighbor and 
next-nearest-neighbor interactions.


\vskip -40mm
\begin{picture}(350,250)
\multiput(60,20)(80,0){5}{\circle{30}}
\multiput(75,20)(80,0){5}{\line(1,0){5}}
\multiput(45,20)(80,0){5}{\line(-1,0){5}}
\multiput(0,20)(80,0){6}{\line(1,-2){5}}
\multiput(35,10)(80,0){6}{\line(1,2){5}}
\multiput(5,10)(80,0){6}{\line(1,4){5}}
\multiput(10,30)(80,0){6}{\line(1,-4){5}}
\multiput(15,10)(80,0){6}{\line(1,4){5}}
\multiput(20,30)(80,0){6}{\line(1,-4){5}}
\multiput(25,10)(80,0){6}{\line(1,4){5}}
\multiput(30,30)(80,0){6}{\line(1,-4){5}}
\put(52,17){\text{n-2}}
\put(132,17){\text{n-1}}
\put(217,17){\text{n}}
\put(290,17){\text{n+1}}
\put(370,17){\text{n+2}}
\put(95,40){\text{$g_2$}}
\put(175,40){\text{$g_2$}}
\put(255,40){\text{$g_2$}}
\put(335,40){\text{$g_2$}}
\qbezier(60,5)(60,-50)(95,-50)
\qbezier(60,5)(60,-50)(25,-50)
\qbezier(220,5)(220,-50)(255,-50)
\qbezier(220,5)(220,-50)(185,-50)
\qbezier(380,5)(380,-50)(415,-50)
\qbezier(380,5)(380,-50)(345,-50)
\multiput(115,-60)(160,0){2}{\line(1,4){5}}
\multiput(120,-40)(160,0){2}{\line(1,-4){5}}
\multiput(125,-60)(160,0){2}{\line(1,4){5}}
\multiput(130,-40)(160,0){2}{\line(1,-4){5}}
\multiput(135,-60)(160,0){2}{\line(1,4){5}}
\multiput(140,-40)(160,0){2}{\line(1,-4){5}}
\multiput(145,-60)(160,0){2}{\line(1,4){5}}
\multiput(150,-40)(160,0){2}{\line(1,-4){5}}
\multiput(155,-60)(160,0){2}{\line(1,4){5}}
\multiput(160,-40)(160,0){2}{\line(1,-4){5}}
\multiput(10,-50)(160,0){3}{\line(1,0){15}}
\multiput(95,-50)(160,0){3}{\line(1,0){15}}
\multiput(110,-50)(160,0){2}{\line(1,-2){5}}
\multiput(165,-60)(160,0){2}{\line(1,2){5}}
\put(135,-35){\text{$g_4$}}
\put(295,-35){\text{$g_4$}}
\qbezier(140,35)(140,90)(175,90)
\qbezier(140,35)(140,90)(105,90)
\qbezier(300,35)(300,90)(335,90)
\qbezier(300,35)(300,90)(265,90)
\multiput(35,80)(160,0){3}{\line(1,4){5}}
\multiput(40,100)(160,0){3}{\line(1,-4){5}}
\multiput(45,80)(160,0){3}{\line(1,4){5}}
\multiput(50,100)(160,0){3}{\line(1,-4){5}}
\multiput(55,80)(160,0){3}{\line(1,4){5}}
\multiput(60,100)(160,0){3}{\line(1,-4){5}}
\multiput(65,80)(160,0){3}{\line(1,4){5}}
\multiput(70,100)(160,0){3}{\line(1,-4){5}}
\multiput(75,80)(160,0){3}{\line(1,4){5}}
\multiput(80,100)(160,0){3}{\line(1,-4){5}}
\multiput(90,90)(160,0){3}{\line(1,0){15}}
\multiput(15,90)(160,0){3}{\line(1,0){15}}
\put(5,90){\line(1,0){10}}
\put(425,90){\line(1,0){10}}
\multiput(30,90)(160,0){3}{\line(1,-2){5}}
\multiput(85,80)(160,0){3}{\line(1,2){5}}
\put(55,110){\text{$g_4$}}
\put(215,110){\text{$g_4$}}
\put(375,110){\text{$g_4$}}
\put(55,43){\text{M}}
\put(135,-10){\text{M}}
\put(215,43){\text{M}}
\put(295,-10){\text{M}}
\put(375,43){\text{M}}
\multiput(60,5)(160,0){3}{\line(0,-1){100}}
\put(140,-85){\vector(1,0){80}}
\put(140,-85){\vector(-1,0){80}}
\put(300,-85){\vector(1,0){80}}
\put(300,-85){\vector(-1,0){80}}
\put(135,-80){\text{$d$}}
\put(295,-80){\text{$d$}}
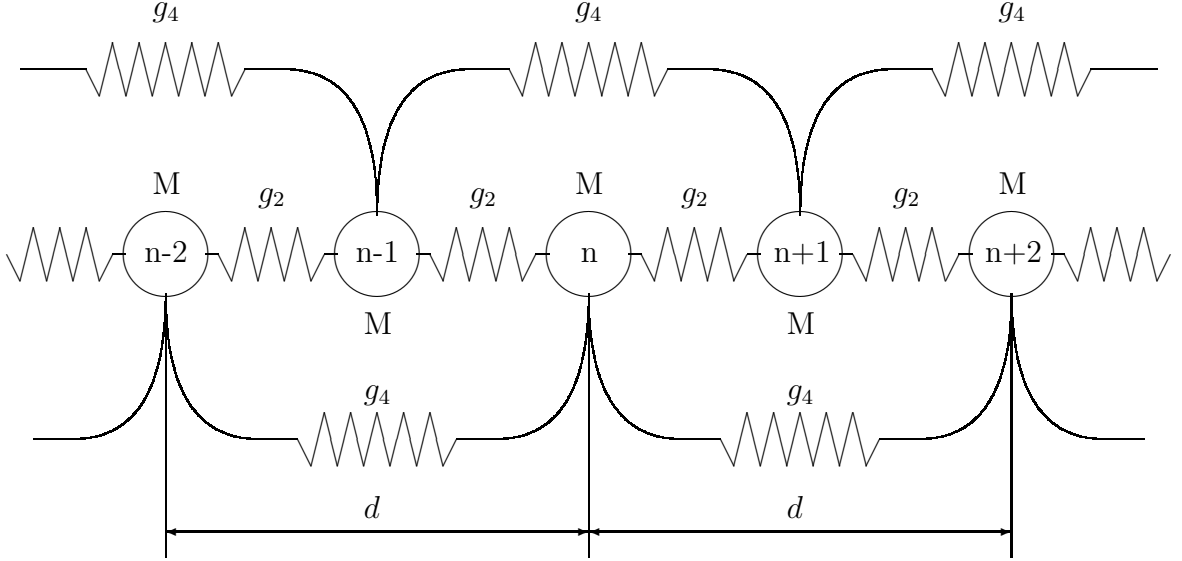
\put(-30,-110){\text{Figure 1: Discrete mass-spring system
with stiffness coefficients $k_2 =g_2$ and $k_4 =g_4$, 
the mass}}
\put(-30,-125){\text{ M and the distance $d$ that correspond to 
the lattice model (\ref{Main_Eq2}) with 
the nearest-neighbor and }} 
\put(-30,-140){\text{next-nearest-neighbor interactions.}}
\end{picture}
\vskip 50mm

\section{Map of lattice in the continuum}

In this paper to derive a continuum equation 
from the lattice model we use the approach suggested 
in \cite{JMP2006,JPA2006} instead of the 
homogenization approach described in Section 2.

In this section, we define the map operation 
\cite{JMP2006,JPA2006,TarasovSpringer} that transforms
the equations of motion for $u_n(t)$ of lattice model 
into continuum equation for a scalar field $u(x,t)$
that describes displacement. 

In order to obtain a continuum equation for a lattice equation, 
we assume that $u_n(t)$ are Fourier coefficients
of some function $\hat{u}(k,t)$.
We define the field $\hat{u}(k,t)$ on $[-K_0 /2, K_0 /2]$ by the equation 
\begin{equation} \label{ukt}
\hat{u}(k,t) = \sum_{n=-\infty}^{+\infty} \; u_n(t) \; e^{-i k x_n} =
{\cal F}_{\Delta} \{u_n(t)\} ,
\end{equation}
where $x_n = n d $ and $d =2\pi/K_0$ is 
distance between equilibrium positions of the lattice particles. 
The inverse Fourier series transform is defined by
\begin{equation} \label{un} 
u_n(t) = \frac{1}{K_0} \int_{-K_0/2}^{+K_0/2} dk \ \hat{u}(k,t) \; e^{i k x_n}= 
{\cal F}^{-1}_{\Delta} \{ \hat{u}(k,t) \} . 
\end{equation}
Equations (\ref{ukt}) and (\ref{un}) are the basis for 
the Fourier transform, which is obtained by transforming 
from lattice variable to a continuum one in 
the limit $d  \to 0$ ($K_0 \to \infty$). 
The Fourier transform can be derived from (\ref{ukt}) and (\ref{un}) 
in the limit as $d  \to 0$.
We replace the lattice function
\[ u_n(t) = \frac{2 \pi}{K_0} u(x_n,t) \] 
with continuous $u(x,t)$ while letting 
\[ x_n=nd = \frac{2 \pi n}{K_0} \to x . \]
Then change ($d \to 0$ or $K_0 \to \infty$) the sum to an integral, and 
equations (\ref{ukt}) and (\ref{un}) become
\begin{equation} \label{ukt2} 
\tilde{u}(k,t)=\int^{+\infty}_{-\infty} dx \ e^{-ikx} u(x,t) = 
{\cal F} \{ u(x,t) \}, 
\end{equation}
\begin{equation} \label{uxt}
u(x,t)=\frac{1}{2\pi} \int^{+\infty}_{-\infty} dk \ e^{ikx} \tilde{u}(k,t) =
 {\cal F}^{-1} \{ \tilde{u}(k,t) \} . 
\end{equation}
We assume that 
\[ \tilde{u}(k,t)= \operatorname{Limit} \hat{u}(k,t), \]
where $\operatorname{Limit}$ denotes the passage 
to the limit $d  \to 0$ ($K_0 \to \infty$).
We note that $\tilde{u}(k,t)$ is a Fourier transform of the field $u(x,t)$,
and $\hat{u}(k,t)$ is a Fourier series transform of $u_n(t)$,
where we can use $u_n(t)=(2\pi/K_0) u(nd ,t)$.
The function $\tilde{u}(k,t)$ can be derived from $\hat{u}(k,t)$
in the limit $d  \to 0$.

As a result, we define the map from a lattice model into a continuum model 
by the following operation \cite{JMP2006,JPA2006}: \\
(1) The Fourier series transform:
\begin{equation} \label{O1}
{\cal F}_{\Delta}: \quad u_n(t) \to {\cal F}_{\Delta}\{ u_n(t)\}=
\hat{u}(k,t) .
\end{equation}
(2) The passage to the limit $d  \to 0$:
\begin{equation} \label{O2}
\operatorname{Limit} : \quad \hat{u}(k,t) \to \operatorname{Limit} \{\hat{u}(k,t)\}=
\tilde{u}(k,t) . \end{equation}
(3) The inverse Fourier transform: 
\begin{equation} \label{O3}
{\cal F}^{-1}: \quad \tilde{u}(k,t) \to 
{\cal F}^{-1} \{ \tilde{u}(k,t)\}=u(x,t) .
\end{equation}

Diagrammatically this set of operations for transformation of 
the displacement can be represented by Figure 2.


\vskip -30mm
\begin{picture}(550,150)
\multiput(60,20)(100,0){4}{\circle{50}}
\multiput(80,23)(100,0){3}{\line(1,0){54}}
\multiput(80,17)(100,0){3}{\line(1,0){54}}
\multiput(140,20)(100,0){3}{\line(-2,1){15}}
\multiput(140,20)(100,0){3}{\line(-2,-1){15}}
\put(45,17){\text{$u_n(t)$}}
\put(145,17){\text{$\hat{u}(k,t)$}}
\put(245,17){\text{$\tilde{u}(k,t)$}}
\put(345,17){\text{$u(x,t)$}}
\put(95,40){\text{${\cal F}_{\Delta}$}}
\put(195,40){\text{$\operatorname{Limit}$}}
\put(295,40){\text{${\cal F}^{-1}$}}
\qbezier(58,0)(60,-52)(95,-52)
\qbezier(62,0)(63,-48)(95,-48)
\qbezier(358,-4)(360,-48)(325,-48)
\qbezier(362,-4)(363,-52)(325,-52)
\put(95,-52){\line(1,0){230}}
\put(95,-48){\line(1,0){230}}
\put(360,0){\line(-1,-2){9}}
\put(360,0){\line(1,-2){9}}
\put(170,-35){\text{$ {\cal F}^{-1} \circ 
\operatorname{Limit} \circ \ {\cal F}_{\Delta}$}}
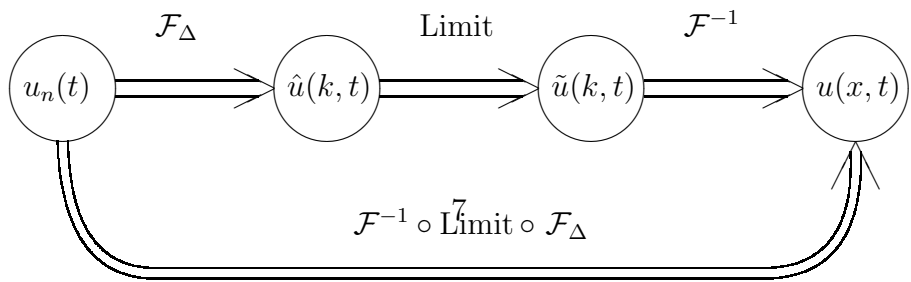
\put(0,-80){\text{Figure 2: Diagrams of 
sets of operations for displacement.}}
\end{picture}
\vskip 30mm

We performed similar transformations for differential equations 
to map the lattice equation into an equation for the elastic continuum.
We can represent these sets of transformations 
of the differential equations in the form of the diagrams
presented by Figure 3.


\begin{picture}(350,250)
\multiput(50,200)(150,0){3}{\oval(110,60)}
\multiput(50,100)(300,0){2}{\oval(130,60)}
\multiput(50,0)(150,0){3}{\oval(110,60)}
\put(20,210){\text{Equation}}
\put(20,190){\text{for $u_n(t)$}}
\put(170,210){\text{From Lattice}}
\put(170,190){\text{to Continuum}}
\put(320,210){\text{Equation}}
\put(320,190){\text{for $u(x,t)$}}
\put(10,110){\text{Fourier series}}
\put(10,90){\text{transform ${\cal F}_{\Delta}$}}
\put(300,110){\text{Inverse Fourier }}
\put(292,90){\text{integral transform  ${\cal F}^{-1}$}}
\put(20,10){\text{Equation}}
\put(20,-10){\text{for $\hat{u}(k,t)$}}
\put(180,10){\text{$\operatorname{Limit}$ }}
\put(180,-10){\text{$d \to 0$}}
\put(320,10){\text{Equation}}
\put(320,-10){\text{for $\tilde{u}(k,t)$}}
\put(45,170){\line(0,-1){40}}
\put(55,170){\line(0,-1){40}}
\put(45,70){\line(0,-1){30}}
\put(55,70){\line(0,-1){30}}
\put(345,160){\line(0,-1){30}}
\put(355,160){\line(0,-1){30}}
\put(345,70){\line(0,-1){40}}
\put(355,70){\line(0,-1){40}}
\put(50,30){\line(-1,2){10}}
\put(50,30){\line(1,2){10}}
\put(350,170){\line(-1,-2){10}}
\put(350,170){\line(1,-2){10}}
\multiput(105,5)(0,200){2}{\line(1,0){40}}
\multiput(105,-5)(0,200){2}{\line(1,0){40}}
\multiput(255,5)(0,200){2}{\line(1,0){30}}
\multiput(255,-5)(0,200){2}{\line(1,0){30}}
\multiput(295,0)(0,200){2}{\line(-2,1){20}}
\multiput(295,0)(0,200){2}{\line(-2,-1){20}}
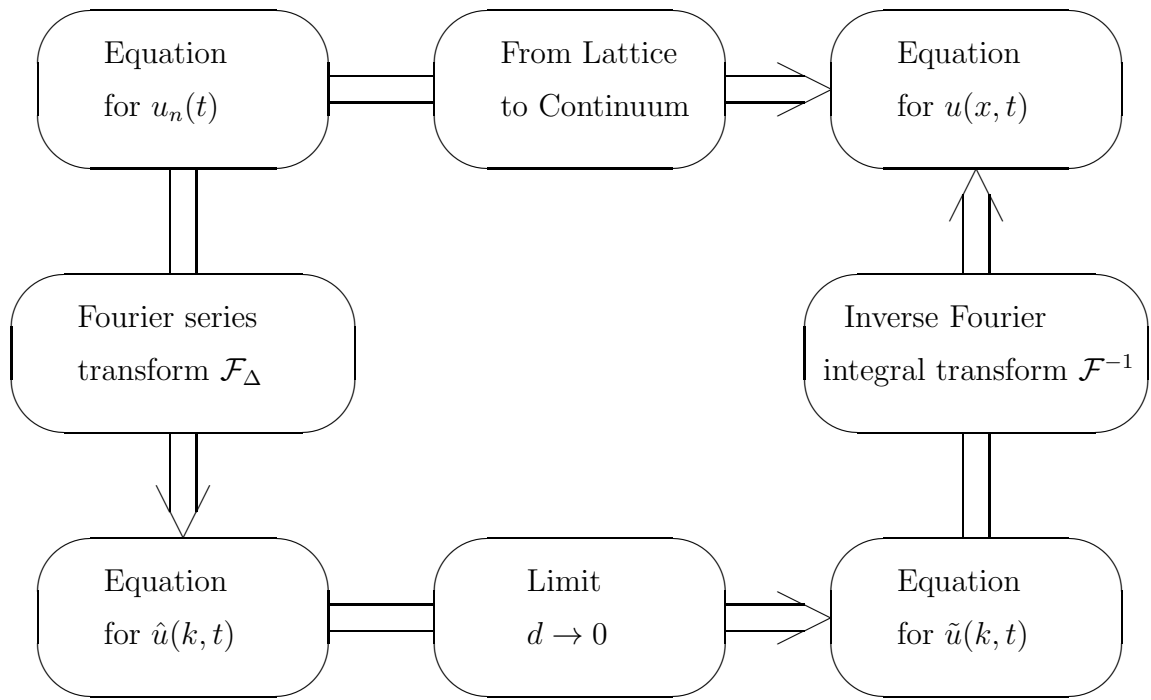
\put(0,-60){\text{Figure 3: Diagrams of 
sets of operations for differential equations.}}
\end{picture}
\vskip 30mm

The combination of these operations ${\cal F}^{-1} \operatorname{Limit} \ {\cal F}_{\Delta}$ 
allows us to realize a map of lattice models of 
interacting particles to models of elastic continuum.

\section{Lattice with nearest-neighbor and next-nearest-neighbor interactions}

Let us consider a lattice with the nearest-neighbor and 
next-nearest-neighbor interactions.
The nearest-neighbor interaction is described by first terms 
on the right-hand side of (\ref{Main_Eq2}) with (\ref{Knm2})
in the form 
\begin{equation}
\sum_{{m=-\infty}}^{+\infty} \; 
K_2(n-m) \; \Bigl( u_n(t)-u_m(t) \Bigl) =
u_{n+1}(t)-2u_n(t)+u_{n-1}(t) .
\end{equation}
The next-nearest-neighbor interaction 
in (\ref{Main_Eq2}) with (\ref{Knm4})
is described by the term 
\begin{equation}
\sum_{{m=-\infty}}^{+\infty} \; 
K_4(n-m) \; \Bigl( u_n(t)-u_m(t) \Bigl) =
u_{n+2}(t)-2u_n(t)+u_{n-2}(t) .
\end{equation}
Here $g_2$ and $g_4$ are coupling constants of the nearest-neighbor and 
next-nearest-neighbor interactions respectively.
In general, we have two different coupling constants.

The corresponding continuum equation can be obtained 
in the limit $d \to 0$ by the method suggested 
in \cite{JMP2006,JPA2006,TarasovSpringer}.
We have the following statement regarding the 
suggested lattice model.

{\bf Proposition.} 
{\it In the continuous limit ($d \to 0$) the lattice equations of motion 
\begin{equation} \label{CEM-2}
M \frac{d^2 u_n}{d t^2} = 
g_2 \cdot \Bigl( u_{n+1}-2u_n+u_{n-1} \Bigr) +
g_4 \cdot \Bigl( u_{n+2}-2u_n+u_{n-2} \Bigr) + F (n) 
\end{equation}
are transformed by the combination ${\cal F}^{-1} \operatorname{Limit} \ {\cal F}_{\Delta}$ 
of the operations (\ref{O1}-\ref{O3}) into the continuum equation: 
\begin{equation} \label{CME0-2}
\frac{\partial^2 u(x,t)}{\partial t^2} = G_2 \, \frac{\partial^2 u(x,t)}{\partial x^2} +
G_4 \, \frac{\partial^4 u(x,t)}{\partial x^4} 
+ \frac{1}{\rho} f(x) ,
\end{equation}
where 
\begin{equation} 
G_2 = \frac{(g_2+4g_4) \, d^2}{M} , \quad 
G_4 = \frac{(g_2+16g_4) \, d^4}{12M} 
\end{equation}
are finite parameters, and $f(x) = F(x)/ (A\, d)$
is the force density, $\rho=M/(A\, d)$ is the mass density. }

{\bf Proof}.
To derive the equation for the field $\hat u(k,t)$, we
multiply equation (\ref{CEM-2}) by $\exp(-ikn d )$, 
and summing over $n$ from $-\infty$ to $+\infty$. Then
\[ \sum^{+\infty}_{n=-\infty} e^{-ikn d } \frac{d^2 u_n}{d t^2} =
g_2 \cdot \sum^{+\infty}_{n=-\infty} \, e^{-ikn d } \, \Bigl( u_{n+1}-2u_n+u_{n-1} \Bigr) + \]
\begin{equation} \label{DD1-b}
+ g_4 \cdot \sum^{+\infty}_{n=-\infty} \, e^{-ikn d } \, \Bigl( u_{n+2}-2u_n+u_{n-2} \Bigr) +
\sum^{+\infty}_{n=-\infty} e^{-iknd } F(n) .
\end{equation}
The first and second terms on the right-hand side of (\ref{DD1-b}) are
\[ g_2 \cdot \sum^{+\infty}_{n=-\infty} \
e^{-ikn d } \, \Bigl( u_{n+1}-2u_n+u_{n-1} \Bigr) +
g_4 \cdot \sum^{+\infty}_{n=-\infty} \
e^{-ikn d } \, \Bigl( u_{n+2}-2u_n+u_{n-2} \Bigr) = \]
\[ = g_2 \cdot \left( \sum^{+\infty}_{n=-\infty} \
e^{-ikn d }  u_{n+1} -
2 \, \sum^{+\infty}_{n=-\infty} \
e^{-ikn d }  u_n +  \sum^{+\infty}_{n=-\infty} \
e^{-ikn d }  u_{n-1} \right) + \]
\[ + g_4 \cdot \left( \sum^{+\infty}_{n=-\infty} \
e^{-ikn d }  u_{n+2} -
2 \, \sum^{+\infty}_{n=-\infty} \
e^{-ikn d }  u_n +  \sum^{+\infty}_{n=-\infty} \
e^{-ikn d }  u_{n-2} \right) = \]
\[ = g_2 \cdot \left( e^{ikd } \, 
\sum^{+\infty}_{m=-\infty} \ e^{-ik m d }  u_{m} -
2 \, \sum^{+\infty}_{n=-\infty} \ e^{-ikn d }  u_n + e^{-ik d } 
\sum^{+\infty}_{j=-\infty} \ e^{-ik j d }  u_{j} \right) +
\]
\begin{equation} \label{Eq-Pr}
= g_4 \cdot \left( e^{2ikd } \, 
\sum^{+\infty}_{m=-\infty} \ e^{-ik m d }  u_{m} -
2 \, \sum^{+\infty}_{n=-\infty} \ e^{-ikn d }  u_n + e^{-2ik d } 
\sum^{+\infty}_{j=-\infty} \ e^{-ik j d }  u_{j} \right) .
\end{equation}
Using the definition of $\hat{u}(k,t)$, equation (\ref{Eq-Pr})
gives
\[ g_2 \cdot \Bigl( e^{ikd } \hat{u}(k,t)- 2 \hat{u}(k,t) +
e^{-ik d } \hat{u}(k,t) \Bigr)+ \]
\[ + g_4 \cdot \Bigl( e^{2ikd } \hat{u}(k,t)- 2 \hat{u}(k,t) +
e^{-2ik d } \hat{u}(k,t) \Bigr)= \]
\[ = g_2 \cdot \Bigl( e^{ikd } +e^{-ik d }-2 \Bigr) \hat{u}(k,t) +
g_4 \cdot \Bigl( e^{2ikd } +e^{-2ik d }-2 \Bigr) \hat{u}(k,t)= 
\]
\[
= 2 \Bigl( g_2 \cdot ( \cos \left( k d  \right)-1 ) + 
g_4 \cdot ( \cos \left(2 k d  \right)-1 )  \Bigr) \hat{u}(k,t) = \]
\[
= 2 \left( 
- 2 \, g_2 \cdot \sin^2 \left( \frac{k d}{2} \right) 
- 8 \, g_4 \cdot \left( \sin^2 \left( \frac{k d}{2} \right) -
\sin^4 \left( \frac{k d}{2} \right) \right)
\right) \hat{u}(k,t) =
\]
\begin{equation} \label{Proof-NNN-1}
= - 4 \, (g_2 +4g_4 )\cdot \sin^2 \left( \frac{k d}{2} \right) \hat{u}(k,t) 
+ 16 \, g_4 \cdot \sin^4 \left( \frac{k d}{2} \right) \hat{u}(k,t) .
\end{equation}
Substitution of (\ref{Proof-NNN-1}) into (\ref{DD1-b}) gives
\begin{equation} \label{simple}
M \frac{\partial^2 \hat{u}(k,t)}{\partial t^2} = 
- 4 \, (g_2 +4g_4 )\cdot \sin^2 \left( \frac{k d}{2} \right) \hat{u}(k,t) 
+ 16 \, g_4 \cdot \sin^4 \left( \frac{k d}{2} \right) \hat{u}(k,t) +
\mathcal{F}_{\Delta} \{ F (n) \} .
\end{equation}
Using the asymptotic behavior of the sine in the form
\begin{equation}
\sin \left( \frac{k d}{2} \right) = \frac{k d}{2} - \frac{1}{6}\left(\frac{k d}{2}\right)^3 + O((kd)^5) ,
\end{equation}
we have
\begin{equation}
\sin^2 \left( \frac{k d}{2} \right) = \frac{(k d)^2}{4} - 2 \, \frac{1}{6} \frac{k d}{2} \frac{(k d)^3}{8} + O((kd)^5) =
\frac{(k d)^2}{4} - \frac{(k d)^4}{48} + O((kd)^5) ,
\end{equation}
\begin{equation}
\sin^4 \left( \frac{k d}{2} \right) = \frac{(k d)^4}{16}  + O((kd)^5) .
\end{equation}
As a result, we can use the representation
\[
- 4 \, (g_2 +4g_4 )\cdot \sin^2 \left( \frac{k d}{2} \right) 
+ 16 \, g_4 \cdot \sin^4 \left( \frac{k d}{2} \right) =
\]
\[
= - 4 \, (g_2 +4g_4 ) \cdot \frac{(k d)^2}{4} 
+ 4 \, (g_2 +4g_4 ) \cdot \frac{(k d)^4}{48} 
+ 16 \, g_4 \cdot  \frac{(k d)^4}{16} =
\]
\begin{equation}
= - (g_2 +4g_4 ) \cdot (k d)^2 + \frac{1}{12} \, (g_2 +16 g_4 ) \cdot (k d)^4 .
\end{equation}
Using the finite parameter $C^2_e=K \, d^2/M$, 
the transition to the limit $d  \to 0$ 
in equation (\ref{simple}) gives 
\begin{equation} \label{DD2-c}
\frac{\partial^2  \tilde u(k,t)}{\partial t^2}=
- G_2 k^2 \tilde u(k,t) + G_4 k^4 \tilde u(k,t) 
+ {\cal F} \{F(x)\} ,
\end{equation}
where 
\begin{equation} \label{Param}
\rho=\frac{M}{A \, d} ,\qquad E= \frac{K \, d}{A}, 
\qquad G_2 =\frac{E}{\rho} = \frac{ (g_2 +4g_4 ) \, d^2}{M} ,
\qquad G_4 = \frac{ (g_2 +16g_4 ) \, d^4}{12M} .
\end{equation}
The inverse Fourier transform ${\cal F}^{-1}$ of (\ref{DD2-c}) has the form
\[ \frac{\partial^2 {\cal F}^{-1}\{ \tilde u(k,t)\} }{\partial t^2}=
- G_2 {\cal F}^{-1} \{k^2 \tilde u(k,t)\} + 
G_4 {\cal F}^{-1} \{k^4 \tilde u(k,t) \}
+ \frac{1}{\rho} f(x) . \]
Then we can use the relation 
${\cal F}^{-1}\{ \tilde u(k,t)\}=u(x,t)$
and the connection between the derivatives and 
its Fourier transforms 
\begin{equation}
{\cal F}^{-1} \{k^2 \tilde u(k,t)\} =
- \frac{\partial^2 u(x,t)}{\partial x^2} , \quad
{\cal F}^{-1} \{k^4 \tilde u(k,t)\} =
+ \frac{\partial^4 u(x,t)}{\partial x^4} .
\end{equation}
As a result, we obtain 
the continuum equation (\ref{CME0-2}). This ends the proof. \\


The correspondent principle and relations (\ref{Param}) gives 
\begin{equation} \label{ggK}
g_2 +4g_4 =K . 
\end{equation} 
It is easy to see that the sign in front the gradient is determined by the value of 
the coupling constant $g_4$ for next-nearest-neighbor interaction.
If we have the inequalities
$g_4 <0$ and $|g_4| > (1/12) \, K $,
then the sign of the constant $G_4$ will be negative.

If we use the kinematic relation 
$\varepsilon (x,t) = \partial u(x,t) /\partial x$, and
the equation of motion of the continuum in the form
\begin{equation} \label{rho-2}
\frac{\partial^2 u(x,t)}{\partial t^2} = 
\frac{1}{\rho} \frac{\partial \sigma(x,t)}{\partial x} +
 \frac{1}{\rho} f(x) ,
\end{equation}
then the constitutive relation is represented as
\begin{equation} \label{sigma1}
\sigma (x,t) = E \, \left( 
\frac{(g_2+4g_4) \, \rho \, d^2}{E \, M} \, \varepsilon (x,t) +
\frac{(g_2+16g_4) \, \rho \, d^4}{12 E \, M} \, \frac{\partial^2 \varepsilon (x,t)}{\partial x^2} \right) 
\end{equation}
with the mass density $\rho= M/ A d$, and 
the Young's modulus $E=K d/A$.
Using (\ref{ggK}), we rewrite relation (\ref{sigma1}) in the from
\begin{equation} \label{sigma2}
\sigma (x,t) = E \, \left( \varepsilon (x,t) + \frac{(K+ 12 g_4) \, d^2}{12 K} \, 
\frac{\partial^2 \varepsilon (x,t)}{\partial x^2} \right) .
\end{equation}
The second-gradient term is preceded by the sign 
that is defined by $\operatorname{sgn} (K+12g_4)$. 
The scale parameter $l^2$ of the gradient elasticity 
is connected with
the coupling constants of the lattice by the equation
\begin{equation} \label{L2}
l^2 =\frac{ \left| K + 12 \, g_4 \right| \, d^2}{12K} .
\end{equation}
Equations (\ref{ggK}) and (\ref{L2}) give 
the close relation between 
the discrete microstructure of lattice and the gradient non-local continuum. 

The sign in front the gradient in (\ref{sigma2}) 
is determined by the relation of the coupling constants 
for nearest-neighbor and next-nearest-neighbor interactions.
We can list all possible cases:

(1) If the coupling constants of the next-nearest-neighbor interaction is
\begin{equation} \label{Negative}
- \frac{1}{4} \, g_2 < g_4 < - \frac{1}{16} \, g_2 ,
\end{equation}
then we have the stress-strain constitutive relation with nagetive sign
\begin{equation}
\sigma (x,t) = E \, \left( \varepsilon (x,t) - l^2 \, 
\frac{\partial^2 \varepsilon (x,t)}{\partial x^2} \right) .
\end{equation}

(2) If we have the inequality
\begin{equation}
g_4 > - \frac{1}{16} \, g_2 ,
\end{equation}
then the sign in the constitutive relation is positive
\begin{equation}
\sigma (x,t) = E \, \left( \varepsilon (x,t) + l^2 \, 
\frac{\partial^2 \varepsilon (x,t)}{\partial x^2} \right) .
\end{equation}

(3) If we have $g_4 < - (1/4) \, g_2$,
then the constant $G_2$ will be negative
and the sign of $G_4$ is the same as one of $g_4$.

(4) If we have $g_4 = - (1/4) \, g_2$,
then the constant $G_2$ and $K$ are equal to zero and 
the sign of $G_4$ is the same as one of $g_4$.

(5) If we have the condition
\begin{equation}
g_4 = - \frac{1}{16} \, g_2 ,
\end{equation}
then the constant $G_4$ is equal to zero and we have 
the well-known Hooke's law without weak nonlocality
\begin{equation}
\sigma (x,t) = E \, \varepsilon (x,t) .
\end{equation}

As a result we can state that the suggested lattice model 
gives us an unified approach to 
the second-gradient elastic continuum models 
with positive and negative signs of the strain gradient terms. 
The close relation between the discrete microstructure of 
lattice and the gradient non-local continuum is given by equations (\ref{ggK}) and (\ref{L2}). 

The proposed lattice model with 
the nearest-neighbor and next-nearest-neighbor interactions 
uniquely leads to second-order strain gradient terms 
that are preceded by the positive and negative signs.
The lattice models with positive value of coupling 
constant $g_4$ of lattice model leads to  
the continuum equation with the positive sign 
in front of the parameter $l^2$. 
This continuum equation is unstable for wave numbers $k> 1/l^2$. 
The instability leads to an unbounded growth
of the response in time without external work. 
The negative value of coupling constant $g_4$ of 
lattice model can lead to stiffness coefficient 
of the next-nearest-neighbor interaction 
with non-convex elastic energy potentials 
in the discrete mass-spring system.
At the same time, the correspondent continuum equation
with the negative sign  in front of the strain gradient
are equivalent to those derived 
from the positive-definite deformation energy density, 
and therefore these continuum models are stable. 
We should note that continuum limits for discrete and lattice systems can be considered without convexity hypotheses
on discrete energy densities \cite{BG2002}.
In addition there exist metamaterials with negative stiffness.
In the plastic deformation range one can observe
a decreasing part of the displacement-force curve, where
stiffness along this part of the curve is negative.
Viscoelastic materials, nanofilms and molecular chains 
containing a negative-stiffness phase, 
anomalies in stiffness and damping have been observed
experimentally \cite{NS1,NS2,NS3,NS5}.
One of the negative stiffness sources can be obtained from 
phase transforming materials in the vicinity of their 
phase transition. 
A theoretical description of the underlying mechanism from 
a microscopic viewpoint has been suggested in \cite{NS6}.
Drugan \cite{NS7} demonstrated that elastic composite materials 
having a negative stiffness phase can be stable.
A stability analysis for elastic composites 
with non-positive-definite phase 
was made by Kochmann and Drugan in \cite{NS8}.
Lee and Goverdovskiy \cite{NS9} demonstrated that
negative-stiffness elements can be used 
to improve vibration isolation systems. 
Elastic composite materials with negative stiffness inclusions
can have extreme overall stiffness and mechanical damping 
\cite{NS10,NS11}.

\section{Conclusion}

In this paper lattice model with the nearest-neighbor and 
next-nearest-neighbor interactions 
for strain-gradient elasticity of continuum is suggested.
This model can be considered as a microstructural basis 
of description for two types of gradient models 
with positive and negative signs.

We can formulate the main results of this paper in 
the following short form:
(a) The method, suggested in \cite{JMP2006,JPA2006} and 
using the Fourier series and integral transformations,
is expanded to elasticity theory;
(b) This method is applied to modified lattice
model with two coupling constants;
(c) The gradient elasticity models with negative sign 
is derived from this lattice model 
that have not be done before.
Let us give some details about the suggested 
lattice model with two coupling constants
and the proposed method that uses the Fourier transformations


In the lattice models of gradient elasticity we suggest
to use two different coupling constants $g_2$ and $g_4$. 
It allows us to derive the constitutive relation 
that contains these constants.
The usual method that based on the direct Taylor expansions   (\ref{Taylor}) allows us to derive 
the constitutive relation (\ref{sign-pos}) 
with positive sign only. 
This traditional approach corresponds to the case $g_4=0$,
where the next-nearest-neighbor interactions 
are not taken into account. 
It cannot give the elastic constitutive 
relations (\ref{H-1n}) with minus sign. 
Only incorporation of two types of interactions 
with two different constants for the nearest-neighbor and 
next-nearest-neighbor interactions may result 
in the continuum limit to the gradient elasticity theory
with positive and negative signs.

The method based on the Fourier series and integral 
transformations is represented by Figure 3. 
This method has been suggested in 
the papers \cite{JMP2006,JPA2006},
and it was applied in \cite{Chaos2006,CNSNS2006}.
Unfortunately this method is not applied 
to models of gradient elasticity.
Moreover there is a common opinion that the gradient models 
with a negative sign in the constitutive relation (\ref{H-1n})  
cannot be obtained from lattice models.
In this paper, the Fourier method is expanded
to lattice model with two coupling constants, 
which is represented by Figure 1. 
It allows us to prove that the gradient continuum 
model with negative sign can be obtained 
from microscopic (lattice) model.
This is one of the main results of this paper. 
We also suggest the conditions (\ref{Negative})
for coupling constants, when the gradient continuum model 
with negative sign can be obtained by the lattice approach. 

Let us note some possible extensions of 
the suggested lattice approach.
Using the second-, third, ... nearest-neighbor interactions 
in the lattice in addition to the nearest-neighbor interaction
it can be easily generalized for the case of 
the high-order gradient elasticity. 
For this generalization, we can use the terms 
$K_{2j}(n,m) = - \Bigl( \delta_{n-m,j} + \delta_{m-n,j}\Bigr)$,
where $j=1,2,3,...$, 
that describe an interaction of the $n$-particle 
with two particles with numbers $n \pm j$.
Using models of lattice with long-range interactions, 
we can get continuum equations with the  
fractional derivatives \cite{KST} of non-integer orders 
by the methods suggested in \cite{JMP2006,JPA2006}.
Therefore the suggested lattice model can be 
extended to get equations for elastic continuum 
with power-law non-locality 
\cite{CEJP2013,MOM2014,ISRN-CMP2014,TA2014,IJSS2014}.
We also assume that the proposed lattice approach
can be generalized for lattice 
with fractal dispersion law \cite{JPA2008,FL-2,FL-3}.
It allows us to get a possible microscopic basis 
for gradient elasticity of fractal materials \cite{TA2015}
that can be described by different tools 
(see for example \cite{CNSNS2015,MOS-7}).
If we consider lattice equation (\ref{Main_Eq}), 
(\ref{Main_Eq2}), (\ref{CEM-2})
with nonlinear site force $F(u_n(t))$ instead of $F (n)$, 
then we can get differential equations 
with nonlinear term $f(u(x,t))$ 
for a continuum in the continuous limit.
It is well-known that the localized soliton solutions 
for lattice models are very important for nonlinear theories
\cite{LatSol-1,LatSol-2}.
Solitons, its interactions, and correspondent 
discrete lattice models can be considered
for different type of physical systems.
For example, equations for an electromagnetic line 
with the second-order couplings can be used 
to describe the bound solitons with oscillating tails 
in this line \cite{GOP}.
The approach suggested in this paper can be generalized
in order to consider continuum models
that correspond to discrete model of a dynamical 
lattice with the on-site nonlinearity and 
both nearest-neighbor and next-nearest-neighbor  
interactions between lattice sites 
\cite{LatSol-3,LatSol-4,LatSol-5}.
We also assume that the unbounded lattice models 
suggested in this paper can be extended to describe 
bounded lattices and correspondent continuum models 
analogous to electromagnetic line described in \cite{GOP}.
We also can assume that suggested approach
can be useful for nonlinear deformable-body 
dynamics \cite{Luo2010} for the case of weak non-locality.


\end{document}